\documentclass[conference]{IEEEtran}
\IEEEoverridecommandlockouts

\usepackage{cite}
\usepackage{amsmath,amssymb,amsfonts}
\usepackage{algorithmic}
\usepackage{graphicx}
\usepackage{textcomp}
\usepackage{xcolor}
\usepackage{url}
\usepackage{booktabs}
\usepackage{array}
\usepackage{hyperref}
\hypersetup{
    colorlinks=true,
    linkcolor=black,
    citecolor=black,
    urlcolor=blue
}

\def\BibTeX{{\rm B\kern-.05em{\sc i\kern-.025em b}\kern-.08em
    T\kern-.1667em\lower.7ex\hbox{E}\kern-.125emX}}

\begin{document}

\title{Heuristic Search Space Partitioning for Low-Latency Multi-Tenant Cloud Queries}

\author{
\IEEEauthorblockN{Prashant Kumar Pathak}
\IEEEauthorblockA{\textit{Palo Alto Networks, Inc.} \\
Santa Clara, CA, USA \\
ppathak@paloaltonetworks.com}
\and
\IEEEauthorblockN{Chandra Biksheswaran Mouleeswaran}
\IEEEauthorblockA{\textit{Palo Alto Networks, Inc.} \\
Santa Clara, CA, USA \\
cmouleeswaran@paloaltonetworks.com}
\and
\IEEEauthorblockN{Rama Teja Repaka}
\IEEEauthorblockA{\textit{Palo Alto Networks, Inc.} \\
Santa Clara, CA, USA \\
rrepaka@paloaltonetworks.com}
}

\maketitle

\begin{abstract}
Large-scale cloud security platforms must continuously query millions of structured cloud resource records distributed across thousands of tenant accounts. Broad, account-spanning queries saturate database infrastructure, producing P95 latencies exceeding 60 seconds. We identify buffer cache pressure as the dominant latency driver: in a controlled experiment, the same query executing with the same plan completed in 3.7 seconds when its working set was memory-resident and 94 seconds when concurrent load had evicted those pages. No query plan optimization can address this; the only effective intervention is reducing the number of pages each query must touch. We present the Heuristic Search Space Partitioning System (HSSPS), a query-time optimization layer that logically partitions the search space through dynamic predicate injection, without schema modification. A two-phase heuristic engine selects partition key values and scores candidate query plans before execution. A client-side page token maintains cross-partition traversal state without server-side sessions, enabling horizontal scalability. Controlled evaluation across representative query types demonstrates 50--97\% P95 latency reduction (95--97\% on high-cardinality queries), 8--10$\times$ throughput improvement, and 41$\times$ reduction in average active sessions. Production rollout across live multi-tenant traffic reduced P95 latency from 61s to 2s across successive releases, sustained over 14{,}000 eligible queries per measurement window. The technique generalizes to any multi-tenant system where broad queries execute against large shared databases and physical schema modification is impractical.
\end{abstract}


\section{Introduction}

Cloud security providers operate platforms that continuously ingest, store, and query structured representations of cloud resources provisioned to customers across major cloud service providers including AWS, Azure, and GCP. A single tenant may operate hundreds of cloud accounts, each containing thousands of provisioned resources. At platform scale---thousands of tenants with concurrent workloads---the resulting databases hold hundreds of millions of resource records that must be queried in real time to support security analysis, compliance monitoring, and threat detection.

The fundamental challenge is inherent to the problem domain: security queries are naturally broad. A query such as ``find all EC2 instances with public IP addresses across all accounts'' carries no natural selectivity predicate and must examine the full tenant resource corpus. For a tenant operating 200 accounts with 50{,}000 resources each, this requires scanning 10 million records per query. At production platform scale, this I/O profile far exceeds standard infrastructure capacity, producing response times that are operationally unacceptable for interactive security workflows.

We observe that conventional optimization approaches---index tuning, query plan optimization, and secondary index addition---do not address the underlying problem. Our measurements show that query latency variability under production load is dominated not by query plan selection but by buffer cache residency. Identical queries with identical plans vary by an order of magnitude in execution time depending solely on whether their working set is memory-resident when executed. Reducing the number of pages a query must touch, not improving how those pages are accessed, is the correct intervention.

This paper presents HSSPS, a system whose central insight is that query-time logical partitioning---augmenting submitted queries with dynamically selected scope predicates---achieves the performance benefits of physical partitioning without any modification to the underlying database structure. HSSPS is transparent to clients, operates uniformly across database engines, and adapts dynamically to live production system state. The technique has been granted U.S. patents US11941006B2 \cite{patent1} and US12373434B2 \cite{patent2} and deployed at production scale.

\subsection{Contributions}

This paper makes the following contributions. First, we identify buffer cache pressure---rather than query plan selection---as the dominant latency driver in multi-tenant cloud resource databases under production load, supported by controlled measurement of identical queries with identical plans exhibiting 25$\times$ latency variance based solely on working set residency. Second, we present a schema-agnostic architecture for query-time search space partitioning using dynamic partition keys, applicable across relational and non-relational databases without schema migration. Third, we describe a two-phase heuristic engine in which partitioning heuristics select partition key values and scoring heuristics evaluate candidate query plans before execution, with the two phases independently tunable. Fourth, we introduce a stateless client-side page token mechanism that maintains cross-partition traversal state without server-side sessions, enabling horizontal scalability. Fifth, we report a controlled evaluation demonstrating 50--97\% P95 latency reduction across representative query types, 8--10$\times$ throughput improvement, and 41$\times$ reduction in average active sessions. Sixth, we report a production rollout validating these results on live multi-tenant traffic: P95 latency reduced from 61 seconds to 2 seconds across successive releases, sustained over a growing population of 14{,}000+ eligible queries per measurement window.

The remainder of this paper is organized as follows. Section~\ref{sec:background} presents background and motivation. Section~\ref{sec:architecture} describes the HSSPS system architecture. Section~\ref{sec:heuristics} details the heuristic engine. Section~\ref{sec:pagination} describes the stateful pagination mechanism. Section~\ref{sec:evaluation} reports our evaluation. Section~\ref{sec:threats} discusses threats to validity. Section~\ref{sec:related} positions HSSPS against related work. Section~\ref{sec:conclusion} concludes.

\section{Background and Motivation}
\label{sec:background}

\subsection{The Cloud Resource Query Problem}

Cloud security platforms ingest resource data from cloud service provider APIs into internal databases. Each record carries structured metadata: resource type, cloud service, account identifier, region, configuration attributes, and ingestion timestamp. The cardinality challenge is fundamental---a query without an account-scoping predicate must execute over the full tenant resource corpus. At a platform serving thousands of concurrent tenants, database I/O demand consistently exceeds provisioned infrastructure capacity, producing latency that routinely exceeds 60 seconds at P95.

This is not a problem that traditional query optimization addresses. The query planner can choose the best plan available, but when that plan must examine tens of millions of pages, execution time is bounded below by I/O capacity, not plan quality. The problem requires an architectural intervention, not a planner improvement.

\subsection{Buffer Cache Pressure as the Dominant Latency Driver}

To establish the dominant mechanism driving latency variability under production load, we executed the same query three times in succession against the same data, within a five-minute window, under typical production concurrent load. The query planner produced an identical execution plan on all three runs: same cost estimate, same access strategy, same operator sequence. Concurrent session counts on the database instance were within $\pm$10\% across the three runs, isolating buffer cache state as the primary differentiating variable. Results are shown in Table~\ref{tab:buffercache}.

\begin{table}[htbp]
\caption{Same Query, Same Plan, Wildly Different Latency}
\label{tab:buffercache}
\centering
\begin{tabular}{@{}lrrr@{}}
\toprule
\textbf{Run} & \textbf{Exec. Time} & \textbf{Buffer Pages} & \textbf{Disk Reads} \\
 & & \textbf{(Shared Hit)} & \\
\midrule
Run 1 & 94 s & 106{,}983 & 21{,}307 pages \\
Run 2 & 3.7 s & 128{,}290 & 0 pages \\
Run 3 & 64 s & 108{,}245 & 20{,}045 pages \\
\bottomrule
\end{tabular}
\end{table}

The difference between 3.7 seconds and 94 seconds is entirely explained by one variable: whether the query's working set resided in the buffer cache. When all 128{,}290 pages were memory-resident, the query completed in 3.7 seconds. When concurrent workloads had evicted those pages, the same query performed 21{,}307 disk reads and took 25$\times$ longer.

No index can eliminate this variability. The query plan was already optimal. The problem is that broad queries touching the entire tenant resource corpus require an enormous number of buffer pages---more than the shared buffer pool can retain under concurrent multi-tenant load. Page eviction is inevitable, and when it occurs, the latency penalty is extreme.

The implication is direct: the only durable intervention is to reduce the number of pages each query must touch. Reducing scan scope, not improving scan efficiency, is the correct mechanism. This insight motivates the HSSPS design.

\subsection{Limitations of Existing Approaches}

Physical sharding \cite{schism} divides a database horizontally by a partition key, enabling shard pruning and reducing buffer pressure per query. Sharding is effective but requires schema migration, is database-engine specific, and on platforms spanning multiple database types demands separate migration paths per engine. On live production systems serving thousands of tenants, this operational risk is often prohibitive.

Secondary indexes improve access paths for selective predicates but do not address the buffer cache mechanism identified above. As our evaluation demonstrates, when a query must touch a large number of pages regardless of access path, index traversal overhead can exceed the cost of a sequential scan---a counter-intuitive degradation observed on high-cardinality queries in our measurements.

Materialized views \cite{matview} precompute results for common patterns but do not generalize to the broad, ad hoc query space characteristic of security analysis. The space of possible queries is too large to pre-materialize exhaustively, and query patterns shift as threat models evolve.

Learned query optimization approaches such as Neo \cite{neo}, Bao \cite{bao}, and learned cardinality estimators \cite{kipf} improve plan selection through machine learning on execution telemetry. These techniques are complementary to HSSPS: they improve the quality of individual query plans, while HSSPS reduces the number of pages those plans must process. In systems deploying both, the effects compound---better plans over smaller search spaces.

Adaptive query processing systems \cite{eddies} adjust execution plans based on runtime observations. HSSPS shares the adaptive philosophy but operates at the query submission layer rather than during execution, enabling schema-agnostic deployment and eliminating the mid-query reoptimization overhead characteristic of runtime adaptation.

HSSPS requires no schema changes, is database-engine agnostic, handles ad hoc queries with no prior knowledge of query patterns, and composes cleanly with both classical and learned optimizations.

\section{System Architecture}
\label{sec:architecture}

HSSPS is positioned between the query interface and the query processor. Clients submit queries in their native language and receive paginated results as if interacting directly with the database. Three principal components compose the system.

The \textbf{Heuristic Query Plan Generator} intercepts submitted queries and determines whether partitioning should be applied. For qualifying queries---those lacking account-level predicates and targeting tables above a configurable cardinality threshold---it generates $N$ candidate augmented queries by appending dynamic partition key predicates.

The \textbf{Heuristic Query Evaluator} applies scoring heuristics to each candidate, evaluating expected execution quality by requesting the execution plan via the database's \texttt{EXPLAIN} facility and extracting estimated row count and execution cost. The highest-scoring candidate is selected for execution.

The \textbf{Stateful Pagination Engine} manages result delivery through a page token mechanism that encodes cross-partition traversal state, enabling complete result assembly across successive requests without any server-side session state.

\subsection{Dynamic Partition Keys}

A dynamic partition key is a designated database field used to logically scope a query to a subset of stored data. In the cloud resource domain, the natural partition key is the cloud account identifier: resource records are natively associated with exactly one account. The key is configurable; HSSPS selects which to apply based on the submitted query and current system metadata. Partition keys may be composite (for example, account combined with region) when the workload distribution requires finer scoping.

\subsection{Query Augmentation}

When HSSPS determines a query should be partitioned, the Plan Generator produces $N$ candidate augmented queries, each appending a scoping predicate of the form:

\begin{center}
\texttt{AND partition\_key IN (v1, v2, ..., vn)}
\end{center}

A concrete example using the cloud account field:

\begin{center}
\small
\texttt{SELECT * FROM resources WHERE cloud.type = 'aws'} \\
\texttt{AND cloud.account IN ('acct\_a', 'acct\_b', ...)}
\end{center}

The augmented query is semantically equivalent to the original restricted to the selected partition key values. By bounding the number of pages each execution must touch, HSSPS directly addresses the buffer cache pressure mechanism identified in Section~\ref{sec:background}. The database is not modified; partitioning is a purely logical, query-time operation fully transparent to the submitting client.

\section{Heuristic Engine}
\label{sec:heuristics}

The heuristic engine operates in two phases. Partitioning heuristics guide the selection of partition key values for candidate queries. Scoring heuristics evaluate and rank those candidates prior to execution. This separation decouples data quality optimization from execution efficiency, enabling independent tuning of each dimension.

\subsection{Partitioning Heuristics}

\textit{Recency heuristic.} Partition key values associated with the most recently updated data are prioritized. Accounts with the most recently ingested or modified resources are surfaced first. Security queries have highest operational value when returning current data; results from inactive accounts are less actionable for ongoing threat detection and compliance.

\textit{Resource count heuristic.} Values with a higher ratio of active to deleted resources are prioritized. Deleted resources inflate estimated row counts and buffer page requirements without contributing to the active result set. Deprioritizing high-deleted-ratio accounts directly reduces buffer cache pressure per execution.

\textit{Relevance heuristic.} For queries referencing specific cloud service APIs or resource types, values are prioritized based on per-account counts of resources matching the queried service. Accounts with zero matching resources produce empty results---wasted I/O and unnecessary buffer eviction. A round-robin algorithm ensures coverage across values over successive executions, preventing starvation of relevant accounts.

\subsection{Scoring Heuristics}

\textit{Relevance score (positive contribution).} Candidates whose partition key values have a higher ratio of active to deleted resources receive a higher composite score, rewarding candidates likely to return denser, more operationally relevant result sets with lower buffer overhead per returned row.

\textit{Cost penalty (negative contribution).} Candidates whose execution plan estimates a larger number of rows to be examined receive a penalty, discouraging executions with large intermediate scan sizes. This directly limits the buffer pages touched per execution, reducing the probability of cache eviction under concurrent workload.

The composite score for a candidate query is the relevance score minus the cost penalty, weighted by tunable coefficients. The candidate with the highest composite score is selected for execution.

\section{Stateful Pagination}
\label{sec:pagination}

\subsection{Page Token Design}

A page token is an opaque token returned to the client with each page of results. It encodes the set of partition key values searched in all previous executions. When the client requests the next page, HSSPS decodes the token, reconstructs the already-searched value set, and excludes these from the next round of heuristic partition selection. All traversal state is carried by the client, eliminating server-side session state entirely and enabling horizontal scalability---any server in the fleet can handle any page request. Clients may stop pagination at any point and receive a consistent partial result set.

\subsection{Security Considerations}

Page tokens are signed using a server-side key to prevent forgery and replay attacks. Tokens encode the originating tenant identifier and carry an expiration timestamp. Tokens presented after expiration or for a mismatched tenant are rejected. Opaque encoding preserves schema opacity across tenant boundaries, preventing token inspection from revealing internal partitioning structure.

\subsection{Termination Criteria}

Execution terminates when one of three conditions holds: all possible partition key values have been searched; a configurable threshold of consecutive empty-result executions is reached; or the client terminates without requesting further pages. The empty-result threshold is separately tunable per query type, allowing aggressive early termination on queries expected to return sparse results and conservative exhaustion on queries requiring completeness.

\section{Evaluation}
\label{sec:evaluation}

We evaluate HSSPS through two complementary methods: a controlled performance evaluation using a representative benchmark workload, and a production rollout validation using live multi-tenant traffic across four successive release boundaries.

\subsection{Deployment Configuration}

The production deployment uses the cloud account identifier as the primary dynamic partition key. The Query Plan Generator produces five candidate augmented queries per partitioning event, each selecting ten partition key values. Partitioning is triggered for queries that do not specify account-level predicates and target tables exceeding a configurable cardinality threshold. Parameter values were determined empirically, balancing plan generation overhead against scan size reduction.

\subsection{Evaluation Methodology}

We evaluate across a workload of thirteen representative query types covering join-heavy and search-heavy access patterns, executed against a multi-tenant cloud resource database spanning hundreds of cloud accounts. Three execution conditions are compared: \emph{unpaginated} baseline with no search space partitioning, \emph{index-assisted} execution using a standard secondary index, and \emph{HSSPS} with heuristic partitioning and stateful pagination enabled. Resource consumption is measured using database Average Active Sessions (AAS) and cache block read rates. Throughput is measured as queries processed per minute under sustained concurrent load.

\subsection{Performance Results}

\begin{table}[htbp]
\caption{HSSPS Performance Improvements Over Unpaginated Baseline}
\label{tab:perf}
\centering
\small
\begin{tabular}{@{}p{2.3cm}p{1.7cm}p{1.3cm}p{1.7cm}@{}}
\toprule
\textbf{Metric} & \textbf{No Part.} & \textbf{HSSPS} & \textbf{Improvement} \\
\midrule
P95 latency (high-cardinality) & 148{,}000--195{,}000 ms & $<$5{,}000 ms & 95--97\% reduction \\
\addlinespace
P95 latency (workload) & 10{,}000--195{,}000 ms & $<$5{,}000 ms & 50--97\% reduction \\
\addlinespace
Peak throughput & 100--150 q/min & 1{,}300 q/min & 8--10$\times$ \\
\addlinespace
Avg active sessions & 16.5 AAS & 0.4 AAS & 41$\times$ reduction \\
\addlinespace
Queries / window & 1{,}800 & 8{,}500 & 4.7$\times$ \\
\addlinespace
Infrastructure cost & Baseline & $-50\%$ & 2$\times$ efficiency \\
\bottomrule
\end{tabular}
\end{table}

\begin{figure}[htbp]
\centerline{\includegraphics[width=\columnwidth]{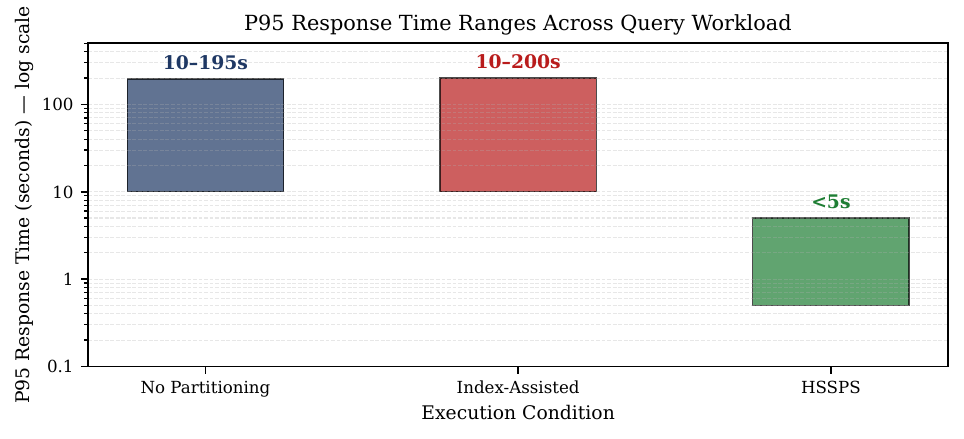}}
\caption{P95 response time ranges across the evaluated query workload (log scale). HSSPS consistently stays under 5 seconds. Index-assisted execution degrades on high-cardinality queries, exceeding the unpaginated baseline---consistent with buffer cache amplification under page eviction pressure.}
\label{fig:latency}
\end{figure}

\textit{Query latency.} Without partitioning, response times range from approximately 10{,}000 ms to 195{,}000 ms across the thirteen query types, with high-cardinality join queries reaching 148{,}000 ms and search-heavy queries reaching 195{,}000 ms. Index-assisted execution degrades on high-cardinality queries, reaching approximately 200{,}000 ms---worse than the unoptimized baseline. This is consistent with the buffer cache analysis in Section~\ref{sec:background}: for large-corpus scans, index traversal overhead compounds page eviction pressure rather than alleviating it. HSSPS execution is flat across all query types, consistently below 5{,}000 ms, representing a 95--97\% latency reduction on high-cardinality queries and 50--97\% reduction across the complete workload. Figure~\ref{fig:latency} summarizes the ranges. The variation reflects differences in how much scan scope HSSPS can prune for each query type; queries with narrower natural cardinality benefit less because the unoptimized baseline is already less saturated.

\begin{figure}[htbp]
\centerline{\includegraphics[width=\columnwidth]{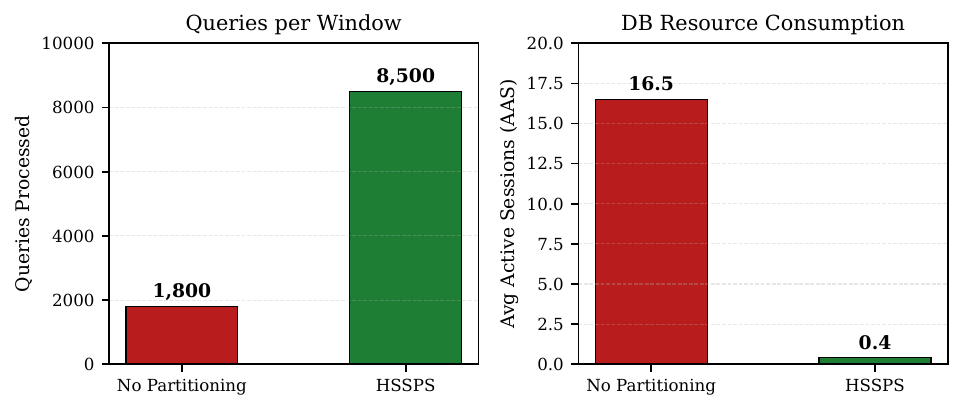}}
\caption{Database resource comparison. HSSPS processes 4.7$\times$ more queries while maintaining 0.4 Average Active Sessions versus 16.5 AAS under unpaginated execution.}
\label{fig:resources}
\end{figure}

\textit{Database resource consumption.} The HSSPS-enabled measurement window processed 8{,}500 queries with AAS measured at 0.4 and flat cache block read rates. The no-partitioning window processed only 1{,}800 queries---4.7$\times$ fewer---while AAS reached 16.5 concurrent sessions with heavy CPU and buffer I/O activity. The 41$\times$ reduction in AAS under HSSPS directly reflects reduced buffer page demand per execution: smaller scan sizes mean fewer pages loaded, lower eviction pressure, and minimal I/O contention across concurrent tenant workloads. Figure~\ref{fig:resources} contrasts the two conditions.

\textit{Throughput.} HSSPS sustains up to 1{,}300 queries per minute during peak load versus 100--150 queries per minute without partitioning---an 8--10$\times$ improvement. The throughput gain is superlinear with respect to the latency reduction, reflecting reduced queuing and buffer contention effects as per-query page demand decreases.

\subsection{Operational Experience}

\textit{Heuristic parameter sensitivity.} The default configuration of five candidates with ten values per candidate performs well across typical tenant distributions. Tenants with atypical account structures---many small accounts versus few large ones---benefit from per-tenant parameter overrides, which HSSPS supports without schema changes because parameters are carried in the query-time partitioning logic rather than in database configuration.

\textit{Resource count cache staleness.} Per-account, per-API resource counts are pre-computed and cached. A 15-minute refresh interval proved sufficient for typical production ingestion rates. Shorter intervals are available for high-churn tenants at the cost of increased metadata pipeline load.

\textit{Empty partition detection.} Executions against newly onboarded accounts may return empty results before ingestion completes. The threshold-based termination criterion handles this gracefully. As a secondary benefit, elevated empty-result rates serve as a real-time signal for detecting ingestion pipeline degradation.

\textit{Cold-start behavior.} For newly provisioned tenants, the recency heuristic falls back to round-robin selection until sufficient resource count history has accumulated. Full heuristic effectiveness typically begins within 24 to 48 hours of initial production traffic.

\subsection{Production Rollout}

To validate that improvements generalize to real workloads, HSSPS was deployed in a phased rollout serving live multi-tenant production traffic, with latency measured directly from production telemetry across four successive release boundaries. HSSPS v1.0 rolled out to a limited tenant cohort to validate production behavior at reduced blast radius; v1.1 expanded to the eligible tenant population; v1.2 extended HSSPS to additional query types beyond the initial scope. Results are shown in Table~\ref{tab:rollout}.

\begin{table}[htbp]
\caption{Production Rollout Latency Across Releases}
\label{tab:rollout}
\centering
\begin{tabular}{@{}lrrr@{}}
\toprule
\textbf{Release} & \textbf{P95} & \textbf{Avg} & \textbf{Eligible} \\
 & \textbf{Latency} & \textbf{Latency} & \textbf{Queries} \\
\midrule
Baseline (Pre-HSSPS) & 61 s & 11 s & 5{,}644 \\
HSSPS v1.0 (limited) & 7 s & 2 s & 1{,}130 \\
HSSPS v1.1 (expanded) & 2 s & 1 s & 10{,}747 \\
HSSPS v1.2 (full) & 5 s & 3 s & 14{,}200 \\
\bottomrule
\end{tabular}
\end{table}

\begin{figure}[htbp]
\centerline{\includegraphics[width=\columnwidth]{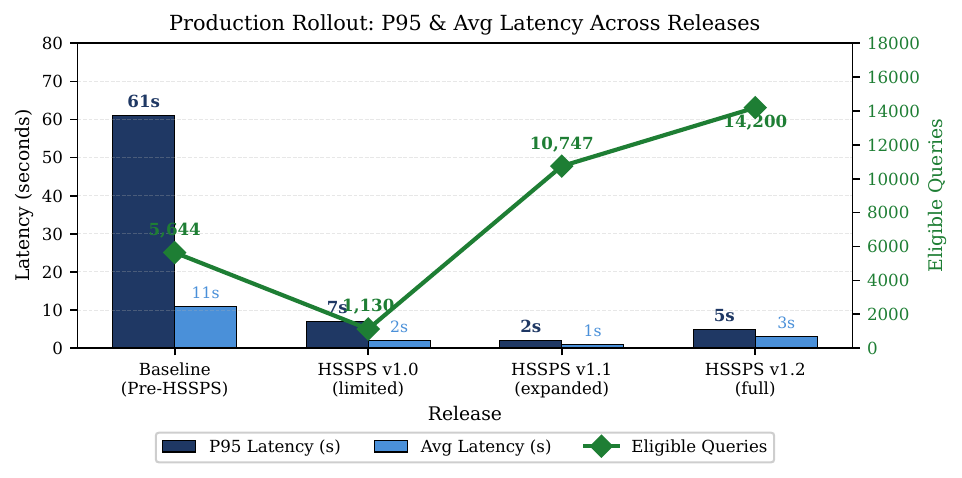}}
\caption{Production rollout results. P95 and average latency decrease from 61s to 2s across releases while eligible query volume grows from 5{,}644 to 14{,}200.}
\label{fig:rollout}
\end{figure}

At HSSPS v1.0, P95 latency decreased from 61 seconds to 7 seconds (88\% reduction) and average latency from 11 seconds to 2 seconds (82\% reduction), consistent with controlled evaluation results. At v1.1, P95 reached 2 seconds against a growing eligible query population of 10{,}747, confirming durability under expanded load. The v1.2 result reflects continued scale-out to additional query types; latency at 5 seconds remains well within interactive response bounds and reflects that newly onboarded query types initially operate with less-mature heuristic calibration. Figure~\ref{fig:rollout} plots both latency trajectories against eligible query volume.

\section{Threats to Validity}
\label{sec:threats}

\textit{Internal validity.} The controlled evaluation covers thirteen query types representative of cloud security workloads, spanning join-heavy and search-heavy access patterns. Generalization to query types with substantially different structures is not directly predicted by the benchmark. The buffer cache variability experiment in Section~\ref{sec:background} was conducted under specific production load conditions with concurrent session counts controlled within $\pm$10\%; variance under substantially different concurrency levels was not systematically characterized and could alter the quantitative results, though the qualitative mechanism---page eviction under load---is well-established in the database literature.

\textit{External validity.} The production evaluation was conducted on one platform in the cloud security domain. Generalizability to other domains---observability, billing, general-purpose multi-tenant SaaS---depends on the similarity of query distributions and tenant cardinality profiles. The core mechanism (logical partitioning to reduce buffer pressure) applies broadly; the specific heuristics may require domain-specific retuning.

\textit{Heuristic calibration and metadata freshness.} Parameter values were determined empirically, and heuristic quality degrades with metadata cache staleness. The 15-minute refresh interval used in evaluation is appropriate for typical production ingestion rates; environments with substantially higher churn may require tighter refresh schedules to maintain equivalent heuristic quality.

\section{Related Work}
\label{sec:related}

Query optimization originates with Selinger et al. \cite{selinger} on cost-based plan selection. Traditional approaches optimize plans through join reordering, predicate pushdown, and index selection. HSSPS operates at the query interface layer and is fully complementary to engine-level optimizations: it reduces the search space before the engine begins, and any engine-level optimization applies independently to the reduced space.

Horizontal partitioning and sharding \cite{schism} physically divide tables by a partition key, enabling shard pruning and reducing buffer pressure per query. Curino et al.'s Schism \cite{schism} introduced workload-driven partitioning as a design principle; HSSPS extends this principle to query-time operation without schema changes. Adaptive systems such as Eddies \cite{eddies} adjust execution plans during execution based on observed runtime statistics; HSSPS shares the adaptive philosophy but operates at the query submission layer, enabling schema-agnostic deployment and eliminating mid-query reoptimization overhead.

Learned query optimization has become an active research area. Marcus et al.'s Neo \cite{neo} and Bao \cite{bao} apply reinforcement learning to plan selection, while learned cardinality estimators \cite{kipf} improve the statistics inputs that guide plan choice. These techniques operate within a fixed search space; HSSPS is complementary, reducing the search space before either classical or learned optimizers begin. In deployments using both, the effects compound.

Multi-tenant database performance has been studied from the isolation angle \cite{multitenant}, with work on resource quotas and per-tenant caching. HSSPS addresses a different dimension: efficient execution of broad queries against the full corpus, which isolation mechanisms do not speed up and may in fact slow down through quota contention. Recent work on cloud-native database performance under multi-tenant load \cite{aurora} has emphasized buffer pool management; HSSPS reduces the load the buffer pool must service, a complementary intervention.

Distributed SQL systems such as Spanner \cite{spanner} and F1 \cite{f1} treat physical partitioning as a first-class architectural concern, with partition-aware query routing built into the system. HSSPS achieves equivalent logical effects for systems where physical partitioning is impractical, making the benefits of partition-aware execution available to workloads running on conventional multi-tenant database deployments.

Cache-aware query processing in multi-tenant environments has been recognized as challenging because shared buffer pools create inter-tenant interference that static optimization cannot address \cite{aurora}. HSSPS directly targets this interference by reducing per-query page demand, and does so without requiring buffer pool partitioning or per-tenant reservation.

\section{Conclusion}
\label{sec:conclusion}

We presented HSSPS, a system that addresses the fundamental challenge of executing broad queries efficiently against large multi-tenant cloud resource databases. Our root cause analysis identified buffer cache pressure as the dominant latency driver---demonstrated by the same query executing in 3.7 seconds or 94 seconds depending solely on working set residency---establishing that reducing pages touched per query, not improving access paths, is the correct intervention.

By intercepting queries at the interface layer and augmenting them with dynamically selected partition predicates, HSSPS achieves the performance benefits of physical partitioning without schema changes, operating uniformly across database engines and adapting dynamically to production state. The two-phase heuristic engine balances result quality against execution efficiency. The stateful page token mechanism enables complete result set traversal without server-side session state.

Controlled evaluation demonstrates 50--97\% latency reduction, 8--10$\times$ throughput improvement, 41$\times$ reduction in average active sessions, and 4.7$\times$ more queries processed per unit of infrastructure. Production rollout confirms these results at scale: P95 from 61 seconds to 2 seconds, sustained across a growing query population exceeding 14{,}000 eligible queries per measurement window.

The core insight---query-time logical partitioning decoupled from physical schema design---extends to any multi-tenant system where broad queries must execute against large shared databases and physical schema modification is impractical. We anticipate adaptation to observability platforms, cloud billing systems, and general-purpose multi-tenant SaaS as fruitful directions for applied extension. The technique has been granted U.S. patents US11941006B2 \cite{patent1} and US12373434B2 \cite{patent2}.

\end{document}